\documentclass{iopart}
\usepackage{amssymb}
\usepackage{iopams}
\usepackage[dvips]{graphicx}

\catcode`\@=11
\renewcommand\footnoterule{%
  \kern-3\p@
  \hrule\@width.4\columnwidth
  \kern2.6\p@}
\renewcommand\@makefntext[1]{%
    \parindent 1em\noindent
    \hb@xt@1.8em{\hss$^{\@thefnmark}$)}\hspace{2pt}%
    \footnotesize\rmfamily#1}  
\def\@makefnmark{\hspace{.5pt}\hbox{$^{\@thefnmark}$%
\hspace{-1pt})}} 
\newcommand{\be}[1]{\begin{equation}\label{#1}}
\newcommand{\ee}{\end{equation}}
\newcommand{\ba}[1]{\begin{eqnarray}\label{#1}}
\newcommand{\ea}{\end{eqnarray}}
\newcommand{\rf}[1]{(\ref{#1})}
\newcommand{\nn}{\nonumber}
\newcommand{\diag}{\mbox{\rm diag}\,}

\newcommand{\sign}{ \mbox{\rm sign}\,}

\def\RR{\mathbb{R}}

\def\CC{\mathbb{C}}

\def\ZZ{\mathbb{Z}}

\newcommand{\bc}{\mathbf{c}}
\newcommand{\bd}{\mathbf{d}}

\newcommand{\bA}{\mathbf{A}}

\newcommand{\vspan}{\mbox{\rm span}\!}

\newcommand{\cD}{\mathcal{D}}
\newcommand{\cH}{\mathcal{H}}

\newcommand{\cK}{\mathcal{K}}
\newcommand{\cL}{\mathcal{L}}
\newcommand{\cP}{\mathcal{P}}
\newcommand{\cS}{\mathcal{S}}
\newcommand{\cT}{\mathcal{T}}

\newcommand{\fu}{\mathfrak{u}}
\newcommand{\fv}{\mathfrak{v}}
\newcommand{\fw}{\mathfrak{w}}

\newcommand{\fA}{\mathfrak{A}}
\newcommand{\fB}{\mathfrak{B}}

\begin{document}

\title[Krein space related perturbation theory for MHD $\alpha^2-$dynamos]{Krein space related perturbation theory for MHD $\alpha^2-$dynamos
and  resonant unfolding of diabolical points}

\author{Oleg N. Kirillov}

\address{Moscow State Lomonosov University, Institute of Mechanics, Michurinskii pr. 1, 119192 Moscow, Russia}
\ead{kirillov@imec.msu.ru}

\author{Uwe G\"unther}

\address{Research Center Rossendorf,  POB 510119,
D-01314 Dresden, Germany}
\ead{u.guenther@fz-rossendorf.de}

\begin{abstract} The spectrum of the spherically symmetric $\alpha^2-$dynamo is
studied in the case of idealized boundary conditions. Starting from
the exact analytical solutions of models with constant
$\alpha-$profiles a perturbation theory and a Galerkin technique are
developed in a Krein-space approach. With the help of these tools a
very pronounced $\alpha-$resonance pattern is found in the
deformations of the spectral mesh as well as in the unfolding of the
diabolical points located at the nodes of this mesh. Non-oscillatory
as well as oscillatory dynamo regimes are obtained. A Fourier
component based estimation technique is developed for obtaining the
critical $\alpha-$profiles at which the eigenvalues enter the right
spectral half-plane with non-vanishing imaginary components (at
which overcritical oscillatory dynamo regimes form). Finally,
Fr\'echet derivative (gradient) based methods are developed,
suitable for further numerical investigations of Krein-space related
setups like MHD $\alpha^2-$dynamos or models of $\cP\cT-$symmetric
quantum mechanics.

\end{abstract}

\submitto{\JPA}

\section{Introduction}

The mean field $\alpha^2-$dynamo of magnetohydrodynamics (MHD)
\cite{mhd-book-1,krause-1,mhd-book-3} plays a similarly paradigmatic
role in MHD dynamo theory like the harmonic oscillator in quantum
mechanics. In its kinematic regime this dynamo is described by a
{\em linear} induction equation for the magnetic field. For
spherically symmetric $\alpha-$profiles $\alpha(r)$ the vector of
the magnetic field can be decomposed into poloidal and toroidal
components and expanded in spherical harmonics. After additional
time separation, the induction equation reduces   to a set of
$l-$decoupled boundary eigenvalue problems
\cite{krause-1,GS-jmp1,GSZ-squire}
\be{i1}
\fA_{\alpha} \fu=\lambda \fu,~~\fu(r\searrow 0)=\fu(1)=0
\ee
for matrix differential operators
\be{i2}
\fA_{\alpha}:=\left(\begin{array}{cc}
            -A_l & \alpha(r) \\
                    A_{l,\alpha} & -A_l \\
                  \end{array} \right)
\ee with \cite{GS-jmp1,GSZ-squire}
\ba{i3}
A_l:=-\partial^2_r+\frac{l(l+1)}{r^2}, \nn \\
A_{l,\alpha}:=-\partial_r\alpha(r)
\partial_r+\alpha(r)\frac{l(l+1)}{r^2}=\alpha(r)A_l-\alpha'(r)\partial_r\,.
\ea
The boundary conditions  in \rf{i1} are idealized ones and formally
coincide with those for dynamos in a high conductivity limit of the
dynamo maintaining fluid/plasma \cite{high-conductivity}. We will
restrict our subsequent considerations to this case and assume a
domain
\ba{i4}
\cD(\fA_{\alpha})=\left\{\fu\in \tilde{\cH}=L_2(0,1)\oplus
L_2(0,1)| \ \fu(r\searrow 0)=\fu(1)=0\right\}
\ea
in the Hilbert space $(\tilde{\cH},(.,.))$. The $\alpha-$profile
$\alpha(r)$ is a smooth real function $C^2(0,1)\ni \alpha(r):\
(0,1)\ \to \ \RR $ and plays the role of the potential in dynamo
models.

Due to the fundamental symmetry of its differential expression
\cite{GS-jmp1,GSZ-squire},
\be{i5}
\fA_{\alpha}=J\fA_{\alpha}^\dagger J,\qquad
J=\left(\begin{array}{cc}0 & I\\ I & 0
 \end{array}\right),
\ee
the operator $\fA_{\alpha}$ is a symmetric operator in a Krein space
$(\cK,[.,.])$ \cite{bognar,langer-LNM-948,azizov,L2,LT-1} with
indefinite inner product $[.,.]=(J.,.)$ and for the chosen domain
\rf{i4} it is also selfadjoint in this space
\be{i6}
[\fA_{\alpha} x,y]=[x,\fA_{\alpha}y], \qquad x,y \in \cK\, .
\ee

Below we analyze the spectrum of the operator $\fA_{\alpha}$ in the
vicinity of constant $\alpha-$profiles --- analytically with the
help of a perturbation theory as well as numerically with a Galerkin
approximation. We obtain a pronounced $\alpha-$resonance pattern in
the occurring deformations of the spectral mesh as well as in the
unfolding of the semi-simple (diabolical \cite{berry-3})
degeneration points which form the nodes of this mesh. Additionally,
we develop a Fourier component based estimation technique for
$(l=0)-$models which allows to obtain the critical $\alpha-$profiles
at which the eigenvalues enter the right spectral half-plane with
non-vanishing imaginary components (at which overcritical
oscillatory dynamo regimes form).

\section{Basis properties of the eigenfunctions in case of constant $\alpha-$profiles}

For constant $\alpha-$profiles $\alpha(r)\equiv \alpha_0=$const, \ \
$r\in [0,1)$,  the operator matrix \rf{i2} takes the simple form
\be{i7}
\fA_{\alpha_0}=\left(\begin{array}{cc}
            -1 & 0 \\
                    \alpha_0 & -1 \\
                  \end{array} \right)\otimes A_l
+\left(\begin{array}{cc}
            0 & \alpha_0 \\
                    0 & 0 \\
                  \end{array} \right)\otimes I
\ee
so that the two-component eigenfunctions $\fu(r)$ can be easily
derived with the help of an ansatz
\be{i8}
\fu_n(r)=\left(\begin{array}{c}
            C_1 \\
               C_2 \\
                  \end{array} \right)\otimes u_n(r)\in \CC^2\otimes
                  L_2(0,1)\, ,
\ee
where $C_1,C_2\in \CC$ are constants to be determined and $u_n$
are the eigenfunctions of the operator $A_l$
\be{i9}
A_l u_n=\rho_n u_n,\qquad u(r\searrow 0)=u(r=1)=0\, .
\ee
These eigenfunctions $u_n$ are Riccati-Bessel functions
\cite{abramowitz}
\be{i10}
u_n(r)=N_n r^{1/2}J_{l+\frac 12}(\sqrt{\rho_n}r),\qquad
N_n:=\frac{\sqrt 2 }{ J_{l+\frac 32}(\sqrt{\rho_n})}
\ee
and we ortho-normalized them as\footnote{See \ref{bessel} for
Riccati-Bessel functions and related orthogonality conditions.}
\be{i11}
(u_m,u_n)=\delta_{mn},\qquad \| u_n\|=1\, .
\ee
Accordingly, the spectrum of the operator $A_l$ consists of simple
positive definite eigenvalues $\rho_n>0$ \  --- \  the squares of
 Bessel function zeros
\be{i12}
J_{l+\frac{1}{2}}(\sqrt{\rho_n})=0, \qquad
0<\sqrt{\rho_1}<\sqrt{\rho_2}<\cdots \, .
\ee

Spectrum and eigenvectors of the operator matrix $\fA_{\alpha_0}$
follow from \rf{i7}, \rf{i8} as
\be{i13}
\lambda_n^\pm=\lambda_n^\pm(\alpha_0)=-\rho_n \pm
\alpha_0\sqrt{\rho_n}\in \RR, \quad n\in \ZZ^+
\ee
and
\be{i14}
\fu_n^\pm=\left(
\begin{array}{c}
  1 \\
  \pm \sqrt{\rho_n} \\
\end{array}
\right)u_n\in \RR^2\otimes L_2(0,1)\, ,
\ee
and correspond to Krein space states of positive and negative type
\be{i15}
[\fu_m^\pm,\fu_n^\pm ]=\pm 2 \sqrt{\rho_n}\delta_{mn},\qquad
[\fu_m^\pm,\fu_n^\mp]=0, \qquad \fu_n^\pm\in \cK_\pm\subset \cK\, .
\ee

\begin{figure}
    \begin{center}
    \includegraphics[angle=0, width=0.95\textwidth]{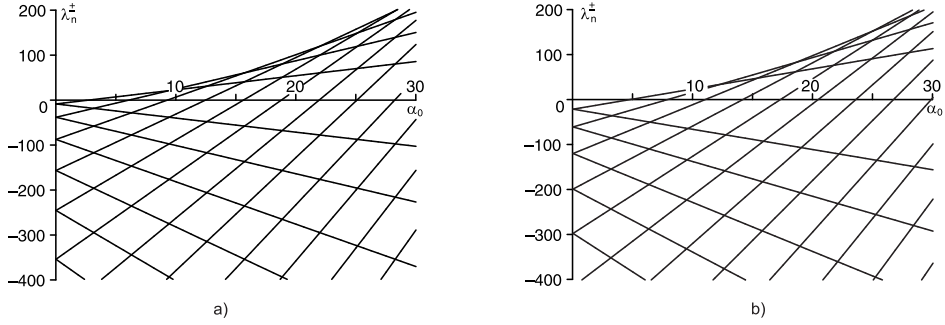}
    \end{center}
    \caption{Eigenvalues $\lambda_{n}^{\pm}(\alpha_0\ge 0)$ for $l=0$ (a) and $l=1$ (b).
    The spectral branches intersect at semi-simple degeneration points (diabolical points)
    of algebraic and geometric multiplicity two.}
    \label{fig1}
    \end{figure}

The branches $\lambda_n^\pm$ of the spectrum are real-valued linear
functions of the parameter  $\alpha_0$ with slopes $\pm
\sqrt{\rho_n}$ and form a mesh-like structure in the $(\alpha_0,\Re
\lambda)-$plane, as depicted in Fig. \ref{fig1}.
In order to calculate the intersection points of the spectral
branches (the nodes of the spectral mesh), we introduce the
following convenient notation
\ba{i16}
\lambda_n^{\varepsilon}&=&-\rho_n+\varepsilon \alpha_0
\sqrt{\rho_n}\, ,\qquad \varepsilon=\pm\nn\\
\fu_n^\varepsilon&=&\left(
\begin{array}{c}
  1 \\
  \varepsilon \sqrt{\rho_n} \\
\end{array}
\right)u_n\, ,
\ea
which allows us to treat positive and negative Krein space states in
a unified way.

Two branches $\lambda_m^\delta,\lambda_n^\varepsilon$ with $n\neq
m$ intersect at a point $(\alpha_0^\nu,\lambda_0^\nu)$ when
\be{i17}
\begin{array}{rcl}
\lambda_n^{\varepsilon}&=&\lambda_m^{\delta}\nn\\
-\rho_n+\varepsilon \alpha_0 \sqrt{\rho_n}&=&-\rho_m+\delta \alpha_0
\sqrt{\rho_m}\nn\\
\alpha_0&=&\frac{\rho_n-\rho_m}{\varepsilon
\sqrt{\rho_n}-\delta   \sqrt{\rho_m}}\nn\\
\alpha_0^\nu :=\alpha_0&=&\varepsilon \sqrt{\rho_n}+\delta
\sqrt{\rho_m}
\end{array}
\ee
and hence
\be{i18}
\lambda_n^{\varepsilon}=\lambda_m^{\delta}=\lambda_0^{\nu
}:=\varepsilon \delta\sqrt{\rho_n\rho_m}\, .
\ee
Eqs. \rf{i17} and \rf{i18} imply that spectral branches of different
type $\delta\neq \varepsilon$ intersect for both signs of $\alpha_0$
at $\lambda_0^\nu <0$. In contrast, intersections at $\lambda_0^\nu
>0$ are induced by spectral branches of positive type when $\alpha_0>0$,
and  by spectral branches of negative type when $\alpha_0<0$.

According to equation \rf{i14} the double eigenvalue
$\lambda_0^{\nu}$ possesses the two distinct eigenvectors
$\fu_n^{\varepsilon}$ and $\fu_m^{\delta}$:
\be{i19}
\fu_n^{\varepsilon}=\left(\begin{array}{c}
            1 \\
            \varepsilon  \sqrt{\rho_{n}} \\
                  \end{array} \right)u_{n},\qquad
\fu_m^{\delta}=\left(\begin{array}{c}
            1 \\
            \delta \sqrt{\rho_{m}} \\
                  \end{array} \right)u_{m}\, .
\ee
Consequently, the intersection points given by \rf{i17} correspond
to double eigenvalues \rf{i18} with two linearly independent
eigenvectors \rf{i19}, i.e. they are semi-simple eigen\-va\-lues or
diabolical points  \cite{berry-3,kato,baumg,KMS05} of algebraic and
geometric multiplicity two.

\section{Unfolding diabolical points by perturbations of the $\alpha$-profile\label{unfold}}
Let us assume that the operator $\fA_{\alpha_0^{\nu}}$ for
$\alpha_0=\alpha_0^{\nu}$ has a semi-simple double eigenvalue
$\lambda_0^{\nu}$ with eigenvectors $\fu_n^{\varepsilon}$ and
$\fu_m^{\delta}$  determined by equations \rf{i18} and \rf{i19}.
Consider a perturbation of the $\alpha$-profile of the form
\be{1p}
\alpha(r)=\alpha_0^{\nu}+\Delta\alpha(r)=\alpha_0^{\nu}+\epsilon
\varphi(r)\, .
\ee
Then, the perturbed operator is given by
\be{2p}
\fA_{\alpha}=\fA_{\alpha_0^\nu}+\epsilon\left(\begin{array}{cc}
            0 & \varphi(r) \\
\varphi(r) A_l - \varphi'(r)\partial_r & 0  \\
                  \end{array} \right)=:\fA_{\alpha_0^{\nu}}+\epsilon\fB
\ee
and the eigenvalue problem can be expanded in terms of the small
parameter $\epsilon$  \cite{kato,baumg,KMS05} as
\be{3p}
(\fA_{\alpha_0^\nu}+\epsilon\fB)(\fu_0^{\nu}+\epsilon \fu_1+\ldots)=
(\lambda_0^{\nu}+\epsilon\lambda_1+\ldots)(\fu_0^{\nu}+\epsilon
\fu_1+\ldots)\, .
\ee
Here, $\fu_0^{\nu}$ is an eigenvector of the unperturbed operator
$\fA_{\alpha_0^\nu}$, corresponds to the eigenvalue
$\lambda_0^{\nu}$ and, hence, has to be a linear combination of
$\fu_n^{\varepsilon}$ and $\fu_m^{\delta}$
\be{4p}
\fu_0^{\nu}=\gamma_1 \fu_n^{\varepsilon}+\gamma_2 \fu_m^{\delta}\in
\vspan\left(\fu_n^{\varepsilon},\fu_m^{\delta}\right)\subset \cK\, .
\ee
A comparison of the coefficients at the same powers of $\epsilon$
yields up to first order in $\epsilon$
\be{5p}
\fA_{\alpha^\nu_0} \fu_0^{\nu}=\lambda_0^{\nu} \fu_0^{\nu}\, ,
\ee
\be{6p}
\fA_{\alpha^\nu_0} \fu_1+\fB \fu_0^{\nu}= \lambda_0^{\nu}
\fu_1+\lambda_1\fu_0^{\nu}\, .
\ee
The first of these equations is satisfied identically, whereas the
second one can be most conveniently analyzed by projecting it with
the help of the Krein space inner product $[.,.]$ onto the
two-dimensional subspace\footnote{The terms containing $\fu_1$
cancel due to the self-adjointness \rf{i6} of the operator
$\fA_{\alpha^\nu_0}$ and Eqs. \rf{4p}, \rf{5p}.} \
$\vspan\left(\fu_n^\varepsilon,\fu_m^\delta\right)\subset \cK$
\be{7p}
[\fB
\fu_0^{\nu},\fu_n^{\varepsilon}]=\lambda_1[\fu_0^{\nu},\fu_n^{\varepsilon}],~~
[\fB
\fu_0^{\nu},\fu_m^{\delta}]=\lambda_1[\fu_0^{\nu},\fu_m^{\delta}]\,.
\ee
Using \rf{4p} in \rf{7p} yields a closed system of defining
equations for the first order spectral perturbation $\lambda_1$
and the coefficients $\gamma_1$ and $\gamma_2$
\be{8p}\fl
\left(%
\begin{array}{cc}
  [\fB \fu_n^{\varepsilon}, \fu_n^{\varepsilon} ] -\lambda_1 [\fu_n^{\varepsilon},\fu_n^{\varepsilon}] & [\fB \fu_m^{\delta}, \fu_n^{\varepsilon}] \\
  ~ & ~ \\
  {[\fB \fu_n^{\varepsilon}, \fu_m^{\delta} ]} & [\fB \fu_m^{\delta}, \fu_m^{\delta}]-\lambda_1[\fu_m^{\delta},\fu_m^{\delta}]
\end{array}%
\right)
\left(%
\begin{array}{c}
  \gamma_1 \\
  ~ \\
  \gamma_2 \\
\end{array}%
\right)=0,
\ee
i.e. in first order approximation the perturbation $\epsilon \fB$
defines the spectral shift $\lambda_1$ and lifts the directional
degeneration \rf{4p} of the zeroth order eigenvectors $\fu^\nu_0$ by
fixing two rays in the subspace
$\vspan\left(\fu_n^\varepsilon,\fu_m^\delta\right)\subset \cK$ (a
standard effect known also from the perturbation theory of
degenerate quantum mechanical systems \cite{landau-3}).

In our subsequent considerations of the system \rf{8p} we will need
different explicit representations of the matrix elements containing
$\fB$. Partial integration and substitution of the relation
$\partial_r^2 u_n=[-\rho_n+l(l+1)/r^2]u_n $ give these
representations as
\ba{10p}
[\fB \fu_m^{\delta}, \fu_n^{\varepsilon}]
&=&\int_0^1\varphi\left[(\rho_m+\varepsilon\delta\sqrt{\rho_n\rho_m})u_mu_n+u_m''u_n+u_m'u_n'\right]dr\label{10p1}\\
&=&\int_0^1\varphi\left[\left((\varepsilon\delta\sqrt{\rho_n\rho_m}+\frac{l(l+1)}{r^2}\right)u_mu_n+u_m'u_n'\right]dr\label{10p2}\,
.
\ea
The symmetry properties of Eq. \rf{10p2} and its implication
\be{10p3} [\fB \fu_m^{\delta},
\fu_n^{\varepsilon}]=[\fB \fu_n^{\varepsilon}, \fu_m^{\delta}]
\ee
are a natural consequence of the Krein space self-adjointness of the
perturbation operator $\fB$ and the real-valuedness of the
eigenvectors $\fu_n^\varepsilon$, $\fu_m^\delta$.

{}From \rf{8p} and \rf{10p3} we obtain the following defining
equation for $\lambda_1$
\be{11p}\fl
\lambda_1^2- \lambda_1\left(\frac{[\fB \fu_n^{\varepsilon},
\fu_n^{\varepsilon} ]}{[\fu_n^{\varepsilon},\fu_n^{\varepsilon}]}+
\frac{[\fB \fu_m^{\delta}, \fu_m^{\delta}
]}{[\fu_m^{\delta},\fu_m^{\delta}]}\right)+\frac{[\fB
\fu_n^{\varepsilon}, \fu_n^{\varepsilon} ][\fB \fu_m^{\delta},
\fu_m^{\delta} ]- [\fB \fu_n^{\varepsilon}, \fu_m^{\delta}
]^2}{[\fu_n^{\varepsilon},\fu_n^{\varepsilon}][\fu_m^{\delta},\fu_m^{\delta}]}=0
\ee
which with the Krein space norm \rf{i15} reduces to
\be{12p}\fl
\lambda_1^2- \lambda_1\left(\varepsilon\frac{[\fB
\fu_n^{\varepsilon}, \fu_n^{\varepsilon} ]}{2\sqrt{\rho_n} }+
\delta\frac{[\fB \fu_m^{\delta}, \fu_m^{\delta} ]}{2\sqrt{\rho_m}
}\right)+\varepsilon\delta\frac{[\fB \fu_n^{\varepsilon},
\fu_n^{\varepsilon} ][\fB \fu_m^{\delta}, \fu_m^{\delta} ] -[\fB
\fu_n^{\varepsilon}, \fu_m^{\delta} ]^2}{4\sqrt{\rho_n\rho_m}}=0\, .
\ee
This quadratic equation is of the type \ $ \lambda_1^2- \lambda_1
(a_1+a_2)+a_1a_2-\varepsilon\delta b^2/4=0,\ \ a_1,a_2,b \in \RR $
and its solutions $\lambda_{1,\pm}=[ (a_1+a_2)\pm
\sqrt{(a_1-a_2)^2+\varepsilon\delta b^2}]/2$ are real-valued for
$\varepsilon =\delta$ and complex for $\varepsilon \neq\delta \
\cap \ (a_1-a_2)^2<b^2$ (in the present first order
approximation\footnote{Higher order corrections may lead to a
further reduction of the real spectral sector.}). We see that they
show the typical Krein space behavior. Intersections of spectral
branches corresponding to Krein-space states of the same type
$(\varepsilon =\delta)$ induce no real-to-complex transitions in
the spectrum (they are weak interactions in the sense of
\cite{sey-maily-book}). In contrast, intersections of spectral
branches corresponding to states of different types $(\varepsilon
\neq\delta)$ may in general be accompanied  by real-to-complex
transitions (they are strong interactions in the sense of
\cite{sey-maily-book}). The same generic behavior is implicitly
present, e.g., in the ${\cal PT}-$symmetric quantum mechanical
(QM) models of Refs.
\cite{PT-diabolic1,PT-diabolic2,PT-diabolic3,PT-diabolic4,PT-diabolic5}
(see also the discussion in \cite{GSG-cz2,GS-cz3}). The unfolding
of a diabolical point in a Hermitian QM model under ${\cal
PT}-$symmetric perturbations was explicitly demonstrated in Ref.
\cite{caliceti-1}.

As noted above, the unfolding of the diabolical points is
accompanied with a fixing of the directions of the zeroth order
eigenvectors $\fu^\nu_0\in
\vspan\left(\fu_n^\varepsilon,\fu_m^\delta\right)\subset \cK$. Using
$\lambda_{1,\pm}$ in \rf{8p} one finds these directions as rays
$\fu^\nu_{0,\pm}=\gamma_{1,\pm}\fu_n^\varepsilon+\gamma_{2,\pm}\fu_m^\delta$
defined by \be{13p}\fl
\frac{\gamma_{1,\pm}}{\gamma_{2,\pm}}=-\frac{[\fB \fu_m^{\delta},
\fu_n^{\varepsilon}]}{[\fB \fu_n^{\varepsilon}, \fu_n^{\varepsilon}
] -
[\fu_n^{\varepsilon},\fu_n^{\varepsilon}]\lambda_{1,\pm}}=-\frac{[\fB
\fu_m^{\delta},
\fu_m^{\delta}]-[\fu_m^{\delta},\fu_m^{\delta}]\lambda_{1,\pm}}{[\fB
\fu_n^{\varepsilon}, \fu_m^{\delta} ]} \ee
--- a generic result obtained, e.g., also in \cite{KMS05}.

As part of the subsequent considerations, we will apply the general
technique \rf{10p1}, \rf{10p2},  \rf{12p} and \rf{13p} for a
detailed analytical study of the unfolding of diabolical points in
concrete dynamo setups.

\section{Local deformations of the spectral mesh\label{deform}}
The perturbation analysis of the previous section has been
restricted to a first order approximation --- giving trustworthy
analytical results for the behavior of the spectrum in a very close
vicinity of any single diabolical point. Here we extend this
approximation method to parameter space regions ($\alpha_0-$regions)
containing several diabolical points --- allowing in this way to
gain a qualitative understanding of how perturbations of the
$\alpha-$profile deform the spectral mesh over such a region. For
this purpose we extend the projection technique of the previous
section from projecting on two-dimensional subspaces
$\vspan\left(\fu_n^\varepsilon,\fu_m^\delta\right)\subset \cK$ to
projections on $N-$dimensional subspaces
$\cL:=\vspan\left(\fu_{n_1}^{\varepsilon_1},\ldots,\fu_{n_N}^{\varepsilon_N}\right)\subset
\cK$ spanned by those eigenvectors which are involved in the
intersections over the concrete region. The method is well known
from computational mathematics as Galerkin method, Rayleigh-Ritz
method or method of weighted residuals
\cite{gottlieb-orszag,bender-orszag,fletcher,boyd,Atkinson}.

In order to simplify notations, we pass from double-indexed
eigenvalues $\lambda_n^\varepsilon$ and states $\fu_n^\varepsilon
\in \cK_\varepsilon \subset \cK_+\oplus \cK_-\subset\cK$ with
$(n,\varepsilon)\in \ZZ^+\times \ZZ_2\sim \ZZ^*=\ZZ-\{0\}$ to
eigenvalues\footnote{We use the notation $\lambda_n$ in spite of a
possible ambiguity in the case of $\lambda_1$. {}From the concrete
context it will be clear whether the first order perturbation
$\lambda_1$ is considered or the spectral branch $\lambda_{n=1}$.}
$\lambda_n$ and normalized states $\fv_n$ depending only on the
single state number\footnote{Depending on the concrete context, we
will subsequently use either double-indexed $\fu_n^\varepsilon$ or
single-index $\fv_n$ notations for convenience.} $n\in \ZZ^*$:
\be{d1}\fl \lambda_n=\left\{\begin{array}{ccc}
            \lambda_n^+ & \mbox{for} & n\in \ZZ^+ \\
            \lambda_{|n|}^- &  \mbox{for} & n\in \ZZ^- \\
           \end{array}\right. , \qquad 2^{1/2}\rho_n^{1/4}\fv_n=\left\{\begin{array}{ccc}
            \fu_n^+ \in \cK_+& \mbox{for} & n\in \ZZ^+ \\
            \fu_{|n|}^- \in \cK_-&  \mbox{for} & n\in \ZZ^- \\
           \end{array}
\right.
\ee
with obvious implication
\be{d2}[\fv_n,\fv_m]= \varepsilon_n\,
\delta_{nm}, \quad \varepsilon_n:=\sign(n)\, .
\ee
Furthermore, we order the index set of the vectors of the subspace
$\cL$ according to the rule $n_1>n_2>\ldots
>n_N$ with $N_+$ vectors $\fv_n$ of positive type and $N_-=N-N_+$ of
negative type.

The approximation of the eigenvalue problem \rf{i1} consists in
representing the eigenfunction $\fu$ as linear combination
\be{d3}
\fu=\sum_{i=k}^N c_k\fv_{n_k},\qquad c_k\in \CC
\ee
over the finite set of basis functions $\{\fv_{n_k}\}_{k=1}^N$ and
projecting\footnote{Given an exact solution $\fu_e$ of the
eigenvalue problem $(\fA_\alpha-\lambda_e)\fu_e=0$, the use of an
approximate test function $\fu=\sum_{i=1}^N c_i\fv_{n_i}$ leads, in
general, to a non-vanishing residual (error)
$R(\bc):=(\fA_\alpha-\lambda_e)\fu$, where
$\bc:=(c_1,\ldots,c_N)^T$. The Galerkin method consists in solving
the eigenvalue problem in a weak sense over the subspace $\cL$
setting $[\fv_{n_i},R(\bc)]=0$ for each of the vectors $\fv_{n_i}\in
\cL$ --- fixing the initially undefined constants $c_i$. This
implies that it yields exact solutions over the subspace $\cL$ and a
non-vanishing residual over the orthogonal complement
$\cL^{[\perp]}=\cK\ominus \cL$. The quality of the approximation can
be naturally increased by increasing the dimension $N=\dim \cL$. A
relatively save test for avoiding spurious solutions in numerical
studies is to compare the output for approximations with different
N. For details on the Galerkin method we refer to \cite{boyd}.} the
resulting equation
\be{d4}
\sum_{i=1}^N c_k\left(\fA_\alpha -\lambda\right)\fv_{n_k}=0
\ee
onto the subspace $\cL=\vspan\,(\fv_{n_1},\ldots,\fv_{n_N})\subset
\cK$. In terms of the notation
\ba{d5}
&&\tilde{\bA}[\alpha]:=(\tilde A_{ij})_{i,j=1}^N,\qquad \tilde
A_{ij}:=\left[\fA_\alpha
\fv_{n_i},\fv_{n_j}\right],\nn\\
&&\bfeta:=\diag(\varepsilon_{n_1},\ldots,\varepsilon_{n_N})=\left(\begin{array}{cc}I_{N_+} & 0\\ 0 & -I_{N_-}\end{array}\right),\nn\\
&&\bc:=(c_1,\ldots,c_N)^T \in \CC^{N_+}\oplus \CC^{N_-}
\ea
this leads to the simple $N\times N-$matrix eigenvalue problem
\be{d6}
(\bA[\alpha]-\lambda I_N )\bc=0, \qquad
\bA[\alpha]:=\bfeta^{-1}\tilde{\bA}[\alpha]
\ee
with
\be{d7}
\det(\bA[\alpha]-\lambda I_N )=0
\ee
as defining equation for the spectral
approximation\footnote{Representing $\alpha(r)$ as function over a
suitably chosen $M-$dimensional parameter space $\cP\ni
(p_1,\ldots,p_M)$, $\alpha(r)=\alpha[r;p_1,\ldots,p_M]$ the
determinant approximation \rf{d7} would allow for easy studies of
the unfolding behavior of the diabolical points over this parameter
space. For example, for $\alpha-$profiles
$\alpha[r;p_1,\ldots,p_M]=\sum_{i=1}^M p_i f_i(r)$ with the
parameters $p_i$ as linear scale factors over a set of test
functions $f_i(r)$ the determinant approximation \rf{d7} will lead
to an algebraic equation $F(\lambda,p_1,\ldots, p_M)=0$ of degree
$\deg (F)=N$ in the spectral parameter $\lambda$ and the parameters
$p_i$. With the help of such an algebraic equation not only the
unfolding of the diabolical points can be tested on their
sensitivity with regard to changes of the functional type $f_i(r)$,
but rather the investigation of other (higher order) types of
algebraic spectral singularities will be easily feasible. For a
discussion  (similar in spirit) on third-order branch points in
$\cP\cT-$symmetric matrix setups we refer to \cite{GS-cz3}.}. A few
comments are in order here.

First, we note that the reality of the eigenvectors $\fv_n$ and of
the dynamo operator $\fA_\alpha$ (cf. \rf{i2} and \rf{i14}) together
with the selfadjointness of $\fA_\alpha$ in the Krein space $\cK$
imply that the matrix $\tilde{ \bA}[\alpha]$ is real and symmetric,
$\tilde{ \bA}^T [\alpha]=\tilde{ \bA}[\alpha]$. The Krein space
related fundamental symmetry $\fA_\alpha=J\fA^\dagger_\alpha J$ (see
\rf{i5})  is reflected in the structure of the matrix
$\bA[\alpha]=\bfeta^{-1}\tilde \bA[\alpha]$ as
\be{d8}
\bA[\alpha]=\bfeta \bA^T[\alpha] \bfeta\,.
\ee
The involutory matrix $\bfeta=\bfeta^{-1}$  plays the role of a
metric in the complex Pontryagin space\footnote{A Krein space
$\cK=\cK_+\oplus\cK_-$ is called a Pontryagin space $\pi_\kappa$
when $\min\left(\dim \cK_+,\dim \cK_-\right)=\kappa<\infty$
\cite{L2}.} $\pi_\kappa=\CC^{N_+}\oplus \CC^{N_-}\ni \bc$. In fact,
it holds for $\fu=\sum_{i=1}^N c_i\fv_{n_i}, \ \fw=\sum_{j=1}^N
d_j\fv_{n_j}$
\be{d9}
[\fu,\fw]=\sum_{i,j=1}^N \bar{c}_i
d_j[\fv_{n_i},\fv_{n_j}]=\sum_{i=1}^N \varepsilon_{n_i}\bar{c}_i
d_i=\bar{\bc}^T\bfeta \bd\,.
\ee

Second, in the limit $N_\pm\to \infty$ the subspace $\cL$ fills the
whole Krein space $\cK$ so that the approximation \rf{d3} of the
vector $\fu$ tends to the exact representation
$\fu=\sum_{n=-\infty}^\infty c_n \fv_n$ over the Krein space basis
$\{\fv_n\}_{n=-\infty}^\infty$. In the same limit, the Pontryagin
space $\pi_\kappa$ tends to the Krein space $\cS=\cS_+\oplus
\cS_-\ni \bc$ with positive and negative type subspaces $\cS_\pm$ as
sequence spaces $\cS_\pm =l_2(\ZZ^\pm)$. The determinant \rf{d7}
becomes a Hill type determinant. The mapping $U$ from the eigenvalue
problem \rf{i1} in the function space $\cK\ni \fu $ to its
equivalent representation \rf{d6} in the sequence space $\cS\ni \bc$
\be{d10}
U:\ \ (\fA_\alpha-\lambda)\fu=0 \quad \mapsto \quad
(\bA[\alpha]-\lambda)\bc=0
\ee
is the Krein space equivalent of the well known mapping from the
quantum mechanical Schr\"odinger picture to its infinite-matrix
representation in the Heisenberg picture \cite{landau-3}. In this
sense, the described Galerkin method can be understood as an
approximate solution technique based on a 'truncated Heisenberg
representation' of the eigenvalue problem. Obviously, the method is
not restricted to $\alpha^2-$dynamo setups, rather in its present
form it is applicable to any other Krein-space related setup as
well, like e.g. models of $\cP\cT-$symmetric quantum mechanics. The
only ingredient needed is a set of exactly known basis functions of
an unperturbed operator.

In the next section, we use the described Galerkin method for a
rough numerical analysis of the deformations of the spectral mesh in
the region depicted in Fig. \ref{fig1} --- leaving analytical
estimates of the residual (the approximation error) to forthcoming
work\footnote{Collaborative work "An operator model for the MHD
$\alpha^2-$dynamo" together with H. Langer and C. Tretter, (in
preparation).}.

\section{The hyper-idealized $s-$wave $(l=0)$ sector\label{l=0}
and its $\bi{\alpha}-$resonance patterns} In this section, we
consider the hyper-idealized case of zero spherical harmonics,
$l=0$, which in analogy to quantum scattering theory can be
interpreted as $s-$wave sector. Due to its too high symmetry
contents, this sector does not play a role in the physics of
spherical dynamos. There are no $s-$wave dynamos at all
\cite{krause-1,GS-jmp1}. Instead, it can be understood as a disk
dynamo model \cite{bar-shu-1} --- in the concrete case of boundary
conditions \rf{i4}, as a disk dynamo with formal boundary conditions
corresponding to a high-conductivity limit. Due to the strong
spectral similarities of models with $l=1$ and $l=0$ (visible e.g.
in Fig. \ref{fig1}), the study of disk dynamo models turns out very
instructive from a technical point of view. Due to their highly
simplified structure they allow for a detailed analytical handling
and a transparent demonstration of some of the essential
mathematical features of the dynamo models. In our concrete context,
they will provide some basic intuitive insight into the dynamo
related specifics of the unfolding of diabolical points. In the
$(l\ge 1)-$sectors of the spherical models,  these specifics will
re-appear in a similar but more complicated way (see section
\ref{l>1} below).

Subsequently, we perform an analytical study of the local unfolding
of the diabolical points (along the lines of Section \ref{unfold})
that we  supplement by numerical Galerkin results on the deformation
of the spectral mesh.

 Let $l=0$. Then the differential expression of the
operator $A_{l=0}$ reads simply $A_{l=0}=-\partial_r^2 $ and
$A_{l=0}$ has ortho-normalized eigenfunctions $u_n(r)$,
$u_n(0)=u_n(1)=0$ and eigenvalues $\rho_n$ \be{h1} u_n=\sqrt 2
\sin(n\pi r), \qquad \rho_n=(\pi n)^2\, . \ee The eigenvalues of the
matrix differential operator $\fA_{\alpha_0}$ are given as \be{h1a}
\lambda_n^\varepsilon=-(\pi n)^2+\varepsilon \alpha_0 \pi n \ee and
the corresponding eigenvectors $\fu_n^\varepsilon$  yield Krein
space inner products \be{h2} \left[ \fu_n^\varepsilon,\fu_m^\delta
\right]=\varepsilon 2\pi n \delta_{\varepsilon \delta}\delta_{mn}
\ee and perturbation terms \be{h3} [\fB \fu_m^{\delta},
\fu_n^{\varepsilon}]=2\pi^2 mn \int_0^1\varphi(r)\cos\left[(\delta
m-\varepsilon n)\pi r\right] dr\, . \ee According to \rf{i17}, the
diabolical points are located at points \be{h3a}
\alpha_0^\nu=\pi(\varepsilon n+\delta m)=:\pi M,\qquad
\lambda_0^\nu=\varepsilon\delta \pi^2 mn \ee and form a periodic
vertical line structure in the $(\alpha_0,\Re
\lambda)-$plane\footnote{In the $(l\ge1)-$sectors (for fixed $l\ge
1$) this line structure is approached asymptotically in the
$|m|,|n|\to\infty$ limit. It follows from substituting the
$l\ll|n|\to\infty$ limit of the Bessel function zeros
\cite{abramowitz}, $\sqrt{\rho_{|n|}}\approx |n|\pi
\left[1+l/(2|n|)\right]$ into the expression for the
$\alpha_0-$coordinate \rf{i17} of the diabolical points:
$\alpha_0^\nu=\varepsilon_n\sqrt{\rho_{|n|}}+\varepsilon_m\sqrt{\rho_{|m|}},
\ \ m,n\in \ZZ^*$.}.

Inspection of the defining equation \rf{11p} for the first order
spectral perturbations $\lambda_1$ shows that this equation is
invariant with regard to a re-scaling of
$\fu_n^\varepsilon,\fu_m^\delta$. Passing to the single-index
notation \rf{d1} yields
\be{h5}\fl \lambda_n=-(\pi
n)^2+\alpha_0\pi n,~~ \fv_n(r)=\frac{1}{\sqrt{\pi
|n|}}\left(\begin{array}{c}
            1 \\
            \pi n \\
          \end{array}
\right) \sin(\pi |n| r),\quad n\in \ZZ^*
\ee
with $[\fv_n,\fv_m]= \varepsilon_n \delta_{nm}$ and the following
convenient representation of the perturbation terms \rf{h3}
\be{h6}
[\fB \fv_{n+j}, \fv_n]=\pi \sqrt{|n(n+j)|}
\int_0^1\varphi(r)\cos(j\pi r) dr\, .
\ee
This reduces the defining equation \rf{12p} for the spectral
perturbation at the $(n,n+j)$ node of the spectral mesh (the
intersection point of the $\lambda_n$ and $\lambda_{n+j}$ branches
of the spectrum) with coordinates
\be{h6a}
\alpha_0^\nu=\pi(2n+j)=:\pi M,\qquad \lambda_0^\nu=\pi^2 n(n+j)
\ee
to
\ba{h7}
&&\lambda_1^2- \lambda_1 \pi (2n+j)\int_0^1\varphi (r) dr
+\nn\\
&&+\pi^2 n(n+j)\left[\left(\int_0^1\varphi (r)
dr\right)^2-\left(\int_0^1 \varphi (r)\cos (j\pi r)
dr\right)^2\right]=0
\ea
with solutions
\ba{h8}\fl
\lambda_{1,\pm}=&\frac \pi 2 (2n+j)\int_0^1\varphi (r) dr \nn \\
\fl & \pm\frac \pi 2 \sqrt{j^2\left(\int_0^1\varphi (r)
dr\right)^2+4n(n+j)\left(\int_0^1\varphi (r) \cos(j\pi r)
dr\right)^2}.
\ea
We observe that the strength of the complex valued unfolding of a
diabolical point at a node with $n(n+j)<0$ is defined by the
relation between its $\cos (j\pi r)-$filtered perturbation $\int_0^1
\varphi (r) \cos(j\pi r) dr$ and its average perturbation $\int_0^1
\varphi (r) dr$.

A deeper insight into this peculiar feature of the unfolding
process can be gained by expanding the perturbation $\varphi$ in
Fourier components over the interval $[0,1]$
\be{h9}
\varphi(r)=\frac{a_0}2 +\sum_{k=1}^\infty\left[a_k \cos(2\pi k
r)+b_k \sin(2\pi k r)\right]
\ee
with coefficients given as $ a_0=2 \int_0^1\varphi(r) dr$, \ \
$a_k=2 \int_0^1\varphi(r) \cos(2\pi k r)dr$, \ \ $b_k=2
\int_0^1\varphi(r) \sin(2\pi k r)dr.$ In this way the integral
$\int_0^1 \varphi(r) \cos(j\pi r) dr$ in \rf{h6} reduces to
components of the type
\ba{h10}
\int_0^1\cos(j\pi r) dr&=&\delta_{j, 0}\, ,\nn\\
\int_0^1\cos(2\pi k r)\cos(j\pi r) dr &=&\frac 12\left(\delta_{j ,
2k} +\delta_{-j , 2k}\right)\, ,\nn\\
\int_0^1\sin(2\pi k r)\cos(j\pi r) dr&=&\left\{
\begin{array}{lcc}
  0& \mbox{\rm for} & j=\pm 2k \\ \\
  \frac {1-(-1)^j}{\pi}\frac{2k}{4k^2-j^2} & \mbox{\rm for} & j\neq\pm 2k
\end{array}\right.
\ea
and we obtain the perturbation terms \rf{h6} as
\ba{h11}
[\fB \fv_{n+j}, \fv_n]=\frac{\pi}{2} \sqrt{|n(n+j)|}Q_{j}\, ,\label{h11a}\\
Q_{j} := a_0\delta_{j,0} +\sum_{k=1}^\infty\left[a_k \left(\delta_{j
, 2k} +\delta_{-j , 2k}\right)+\frac {1-(-1)^j}{\pi}\frac{4b_k
k}{4k^2-j^2}\right].\label{h11b}
\ea
The defining quadratic equation for the spectral perturbation
$\lambda_1$ takes now the form
\be{h12}
\lambda_1^2- \lambda_1
\frac{\pi}{2}(2n+j)a_0+\frac{\pi^2}{4}n(n+j)\left(a_0^2-Q_{j}^2\right)=0
\ee
and leads to the following very instructive representation for its
solutions
\be{h13}
\lambda_{1,\pm}=\frac \pi 4 \left[ (2n+j)a_0\pm \sqrt{j^2 a_0^2
+4n(n+j)Q_{j}^2}\right].
\ee

The structure of these solutions shows that the unfolding of the
diabolical points is controlled by several, partially competing,
effects. Apart from the above mentioned  Krein space related feature
of unfolding into real eigenvalues for states of the same Krein
space type $(n(n+j)>0)$, and the possibility for unfolding into
pairwise complex conjugate eigenvalues in case of states of opposite
type $(n(n+j)<0)$ a competition occurs between oscillating
perturbations $(a_{k\neq 0},b_k)$ and homogeneous offset-shifts
$a_0$. In the case of vanishing offset-shifts (mean perturbations),
$a_0=0$, any inhomogeneous perturbation with $a_k,b_k\neq 0$ for
some $k\ge 1$ leads for two branches with $n(n+j)<0$ to a complex
unfolding of the diabolical point. The strength of this complex
directed unfolding becomes weaker when the homogeneous offset
perturbation is switched on ($a_0\neq 0$). There exists a critical
offset
\be{h14}
a_{0(c)}^2:=-\frac{4n(n+j)}{j^2}Q^2_{j}
\ee
which separates the regions of real-valued and complex-valued
unfoldings\footnote{We note that for intersecting spectral branches
it necessarily holds $j\neq 0$ so that the offset term
$a_0\delta_{j,0}$ in $Q_{j}$ cancels and the definition \rf{h14} of
the critical offset $a_{0(c)}^2$ is justified.} (in the present
first-order approximation). For $ a_0^2<a_{0(c)}^2 \ \cap \ n(n+j)<0
$ a complex-valued unfolding occurs, whereas for $ a_0^2>a_{0(c)}^2
\ \cap \ n(n+j)<0 $ the diabolical point unfolds real-valued. The
special case of a critical (balanced) perturbation $a_0^2=a_{0(c)}^2
\ \cap \ n(n+j)<0$ corresponds to an eigenvalue degeneration
$\lambda_{1,+}=\lambda_{1,-}=:\lambda_{1(c)}$  which, because of
\rf{13p}, \rf{d1} and
$\gamma_{1,+}\gamma_{2,+}^{-1}=\gamma_{1,-}\gamma_{2,-}^{-1}\, $,
has coinciding zeroth order rays so that via appropriate
normalization the corresponding vectors
$\fv^\nu_{0,+},\fv^\nu_{0,-}$ can be made coinciding
$\fv^\nu_{0,+}=\fv^\nu_{0,-}=:\fv^\nu_{0(c)}$. This means that the
original diabolical point splits into a pair of exceptional (branch)
points at perturbation configurations $a_0=\pm |a_{0(c)}|$ with a
Jordan chain consisting of the single (geometric) eigenvector (ray)
$\fv^\nu_{0(c)}$ supplemented by an associated vector (algebraic
eigenvector). This is in agreement with the unfolding scenario of
diabolical points of general-type complex matrices described e.g. in
\cite{KMS05}.

Finally, the special case of $j=-2n$ is of interest. It corresponds
to the intersection points located on the $(\alpha_0=0)-$axis of the
$(\alpha_0,\Re \lambda)-$plane, where the operator matrix is not
only self-adjoint in the Krein space $\cK$, but also in the Hilbert
space $\tilde \cH$. Due to the vanishing factor $2n+j$  these
diabolical points unfold via perturbations $\lambda_1=\pm (\pi/2)
|n|\sqrt{a_0^2- a_{|n|}^2}$.

    \begin{figure}
    \begin{center}
    \includegraphics[angle=0, width=0.55\textwidth]{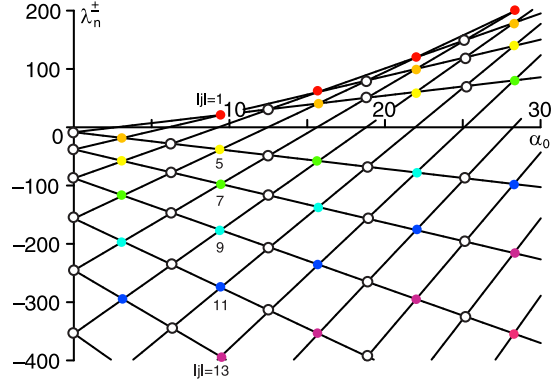}
    \end{center}
    \caption{Odd harmonics of $\alpha(r)$ define the unfolding properties only of the colored diabolical
    points (equally colored DPs correspond to the same value of $|j|$), whereas even harmonics affect
    only the white (uncolored) DPs. }
    \label{fig2}
    \end{figure}

Let us now consider the strength of the unfolding contributions
induced by certain Fourier components. For this purpose, we note
that the diabolical points at nodes $(n,n+j)$ with  the same
absolute value of the index $j$ are located on a parabolic
curve\footnote{These $(\alpha_0,j)-$parametrized parabolic curves
coincide exactly with the spectral curves of a $(l=0)-$sector model
with physically realistic boundary conditions \cite{GK-3}. (For a
discussion of physically realistic boundary conditions of
spherically symmetric $\alpha^2-$dynamos we refer to
\cite{krause-1,GS-jmp1}.)} \be{h16}
\lambda_0^{\nu}=\frac{1}{4}({\alpha_0^{\nu}}^2-\pi^2
j^2)=\frac{\pi^2}{4}(M^2- j^2)\, . \ee Due to its special role, we
will refer to $j\in \ZZ$  as parabola index ($M\in \ZZ$ is the index
of the vertical line in the $(\alpha_0,\Re \lambda)-$plane defined
in \rf{h3a} and \rf{h6a}). Furthermore, we see from the explicit
structure of $Q_{j}$ (following from expression (\ref{h11b}))
\be{h15} Q_j=\left\{\begin{array}{ll}
  \frac{8}{\pi}\sum_{k=1}^{\infty} b_k \frac{k}{4k^2-j^2}\, ,\quad & j=\pm1,\pm3,\ldots \\
  a_k\left(\delta_{j,2k}+\delta_{-j,2k}\right)\, , & j=\pm2,\pm4,\ldots
\end{array}\right.
\ee
that cosine and sine components of a similar order $|a_k|\sim |b_k|$
contribute differently at different nodes of the spectral mesh. It
is remarkable that for all DPs with the same {\it even} $|j/2|$
(parabolas consisting of white points in Fig.~\ref{fig2}), the
splitting of the corresponding double eigenvalues depends (modulo
the pre-factors $4n(n+j)$, $j^2$) only on the mean value $a_0$ of
the perturbation $\varphi(r)$ and its $|j/2|$-th cosine component
$a_{|j/2|}$. For $n(n+j)<0$ these contributions are competing,
whereas for a strictly real-valued unfolding $n(n+j)>0$ they enhance
each other. Furthermore, we find from $M:=2n+j$ that the even/odd
mode properties of $j$ imply the same properties for $M$: even (odd)
modes affect the unfolding of diabolical points at even (odd) $M$
only. This is also clearly visible from  Fig. \ref{fig2}.

In the more complicated case of {\it odd} parabola indices $|j|$
the splitting of the diabolical points on the parabola (\ref{h16})
(in Fig.~\ref{fig2} they are marked as points of the same color)
is governed by the competition between $a_0$ and the complete set
of sine components $b_k$ of the perturbation $\varphi(r)$.
According to (\ref{h15}), a dominant role is played by  sine
harmonics with $j^2 \approx 4k^2$, i.e. with
$k_{\pm}=(|j|\pm1)/2$,  such that a clear resonance and damping
pattern occurs. Sine components with $j^2 \approx 4k^2$ are highly
enhanced by a small denominator over the other sine components and
the corresponding harmonics can be regarded as resonant ones. In
contrast, sine contributions with modes away from the resonant
$k_{\pm}=(|j|\pm1)/2$ are strongly damped by the denominator
$4k^2-j^2$ and tend asymptotically to zero for $j^2\to \infty$.

A first order approximation based insight into this asymptotical
behavior can be gained from the positions of the exceptional points
(EPs) in the $(\alpha_0,\Re \lambda)-$plane (the points where
real-to-complex transitions occur). For this purpose, we use a
1D-lattice type parametrization for $\alpha_0$ in form of
$\alpha_0=\alpha_0^\nu+\Delta\alpha_0=\pi M +\Delta\alpha_0, \
M\in\ZZ$ and switch to a setting with $\epsilon=1$ and
$\Delta\alpha(r):=\phi(r)$. Interpreting $\Delta\alpha_0$ as
perturbation of a configuration with $\alpha_0^\nu=\pi M$ allows us
to relate $\Delta\alpha_0$ to the Fourier coefficient
$a_0=2\Delta\alpha_0$ and to estimate the $\alpha_0-$positions of
the EPs $\Delta\alpha_e(M,j)$ relative to their corresponding
diabolical points located on the line $\alpha_0^\nu=\pi M$. Via
\rf{h14} and $n=(M-j)/2$ we get for given $Q_j$ \ba{h15-1}
\left[\Delta\alpha_e(M,j)\right]^2=-\frac{n(n+j)}{j^2}Q_j^2=\frac
14\left(1-\frac{M ^2}{j^2}\right)Q^2_j\,. \ea For diabolical points
in the lower $(\alpha_0,\Re\lambda)-$half-plane it holds $|j|\ge
|M|+2$ and \rf{h15-1} is well defined. In the case of cosine
perturbations $a_k\cos (2\pi kr)$, \rf{h15} implies that only a
single diabolical point per $M=0,\pm 2,\pm 4,\ldots$ unfolds,
whereas for sine perturbations $b_k\sin(2\pi kr)$ it leads to a
countably infinite number of unfolding diabolical points per $M=\pm
1, \pm 2,\ldots$. Substituting \rf{h15} into \rf{h15-1} one obtains
the $M^2,k^2\ll j^2\to \infty$ asymptotics of the EP positions as
\be{h15-2} |\Delta\alpha_e(M,j)|\approx \frac{4|b_k|k}{\pi j^2}, \ee
i.e. for increasing $j^2$ the distance of the EPs from the DPs is
tending to zero. Conversely, \rf{h15-2} may be used to give an
estimate for the number of complex eigenvalues for a given
$\alpha_0=\pi M+\Delta\alpha_0$ close\footnote{The first-order
approximation \rf{h15-1} leads to a third-order polynomial
$F(\Delta\alpha_e,j^2)=0$ in $j^2$. Its solutions are of limited
meaning because of possible contributions from higher-order
perturbation terms.} to a diabolical point line at $\alpha_0\approx
\pi M$, $|\Delta\alpha_0|\ll \pi$: \ $ j^2\approx 4|b_k|k/(\pi
|\Delta\alpha_0|). $

Above, we arrived at the conclusion that both types of $|j|-$nodes
(even ones and odd ones) show a similar collective behavior along
the parabolic curves (\ref{h16}) of fixed $j^2$, responding on some
specific $\alpha-$perturbation harmonics in a resonant way. Hence, a
specific $\alpha-$resonance pattern is imprinted in the spectral
unfolding picture of the diabolical points.
\begin{figure}[htb]
    \begin{center}
    \includegraphics[angle=0, width=1.0\textwidth]{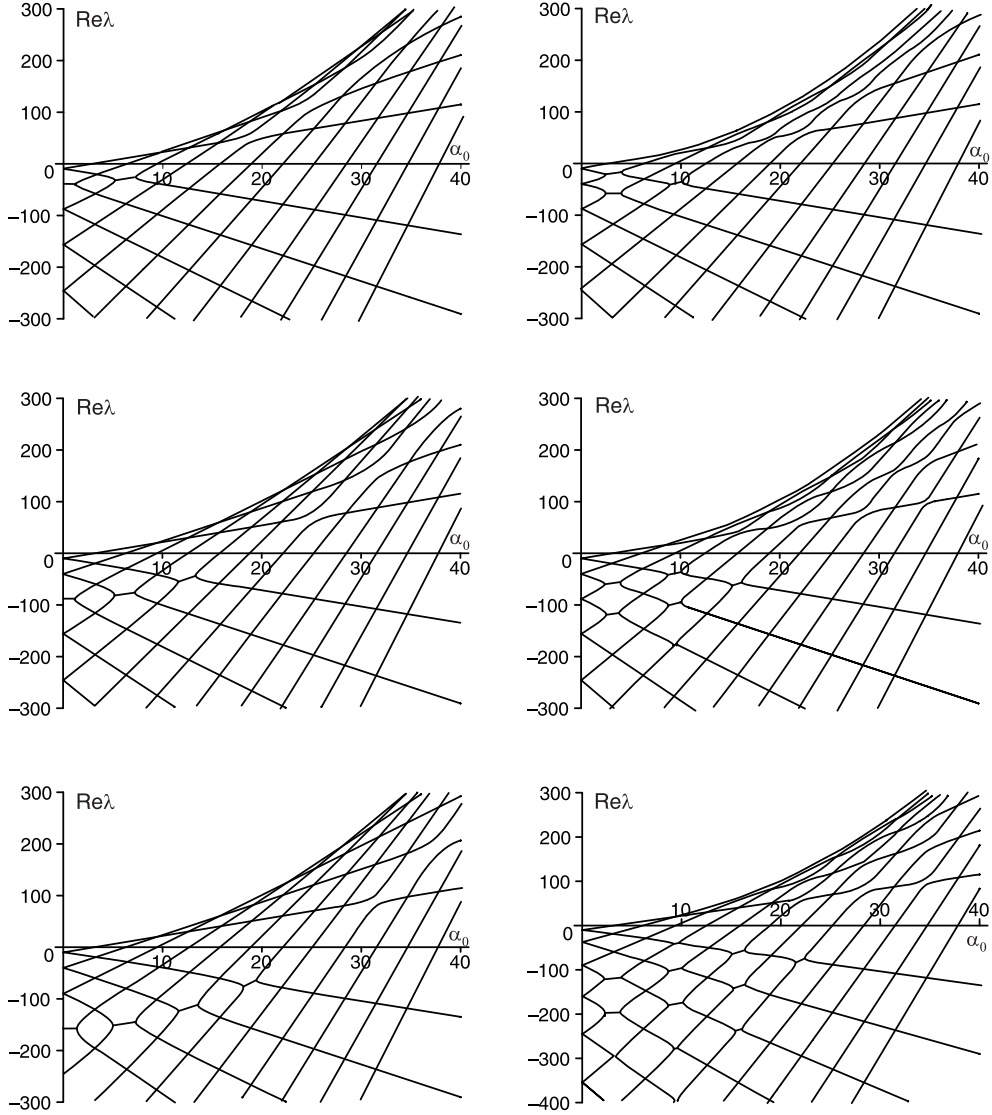}
    \end{center}
    \caption{Resonant deformation of the spectral mesh with resonant unfolding of diabolical points
    due to perturbations by pure harmonics.
    Left column top down $\Delta\alpha(r)=2.5\cos(4\pi r)$,
    $\Delta\alpha(r)=2.5\cos(6\pi r)$, $\Delta\alpha(r)=2.5\cos(8\pi r)$;
    right column top down $\Delta\alpha(r)=2.5\sin(4\pi r)$,
    $\Delta\alpha(r)=2.5\sin(6\pi r)$, $\Delta\alpha(r)=2.5\sin(8\pi r)$.}
    \label{fig3}
    \end{figure}
Let us now analyze these $\alpha-$resonance patterns as imprints in
the deformations of the spectral mesh. We study these deformations
with the help of a Galerkin approximation over a 24-dimensional
Krein subspace $\cL=\vspan(\fv_{12},\ldots, \fv_{-12})\subset \cK$
--- which is sufficient to cover the same spectral region as in Fig. \ref{fig1}.
As $\alpha-$profile we choose $\alpha (r)=\alpha_0+\Delta \alpha
(r)$ with pure harmonics $\Delta \alpha(r)=a_k \cos(2 \pi k r)$ and
$\Delta \alpha(r)=b_k \sin(2 \pi k r)$, $k=2,3,4$, \ $a_k=b_k=5/2$
as perturbations\footnote{The amplitudes $a_k, b_k$ have been chosen
as large as $a_k=b_k=5/2>1$ in order to clearly demonstrate the
unfolding pattern.}. The explicit structure of the corresponding
Pon\-trya\-gin space related matrix $\bA[\alpha]$ (see
 \rf{d6}) is given in \ref{galerkin} and yields the
spectral approximations depicted in Fig.~\ref{fig3}.

The most striking feature of the spectral deformations is their very
clearly pronounced resonance character along parabolas with fixed
index $|j|$ --- leaving spectral regions away from these resonance
parabolas almost unaffected. Specifically, we find for {\em cosine}
perturbations (depicted in the left column) that the harmonics
$k=2,3,4$ affect only the unfolding of diabolical points located
strictly on the associated parabolas with index $j=2k$. The effect
of {\em sine} perturbations with mode numbers $k=2,3,4$ is shown in
the right column graphics of Fig. \ref{fig3}. As predicted by
\rf{h15}, we find a strongly pronounced unfolding of diabolical
points located on the parabolas with $|j|=2k\pm1$, that is for
$|j|=3$ and $|j|=5$ (upper right picture), $|j|=5$ and $|j|=7$
(middle right), and $|j|=7$ and $|j|=9$ (lower right picture). The
DPs with $|j|=2k\pm m$, $m>1$ are less affected and the strength of
the unfolding quickly decreases with increasing distance $m$ to the
resonant parabolas. In addition to the unfolding effects predicted
analytically by first order perturbation theory, the top and middle
right pictures show additional DP unfoldings on the
large$-\alpha_0-$end of the $|j|=2k$ parabola. The origin of these
unfoldings can be attributed to higher-order perturbative
contributions.

The $\alpha-$resonance pattern has a simple physical interpretation.
Due to the fact that the spectral parameter implies an $e^{\lambda
t}$ behavior of the corresponding field mode, the $\alpha-$resonance
pattern shows that short scale perturbations of the $\alpha-$profile
(Fourier components with higher $k$) coherently affect faster
decaying (more negative $\Re \lambda$) field modes than large scale
perturbations with smaller $k$ (which lead to smaller negative $\Re
\lambda$).

\begin{figure}[htb]
    \begin{center}
    \includegraphics[angle=0, width=0.9\textwidth]{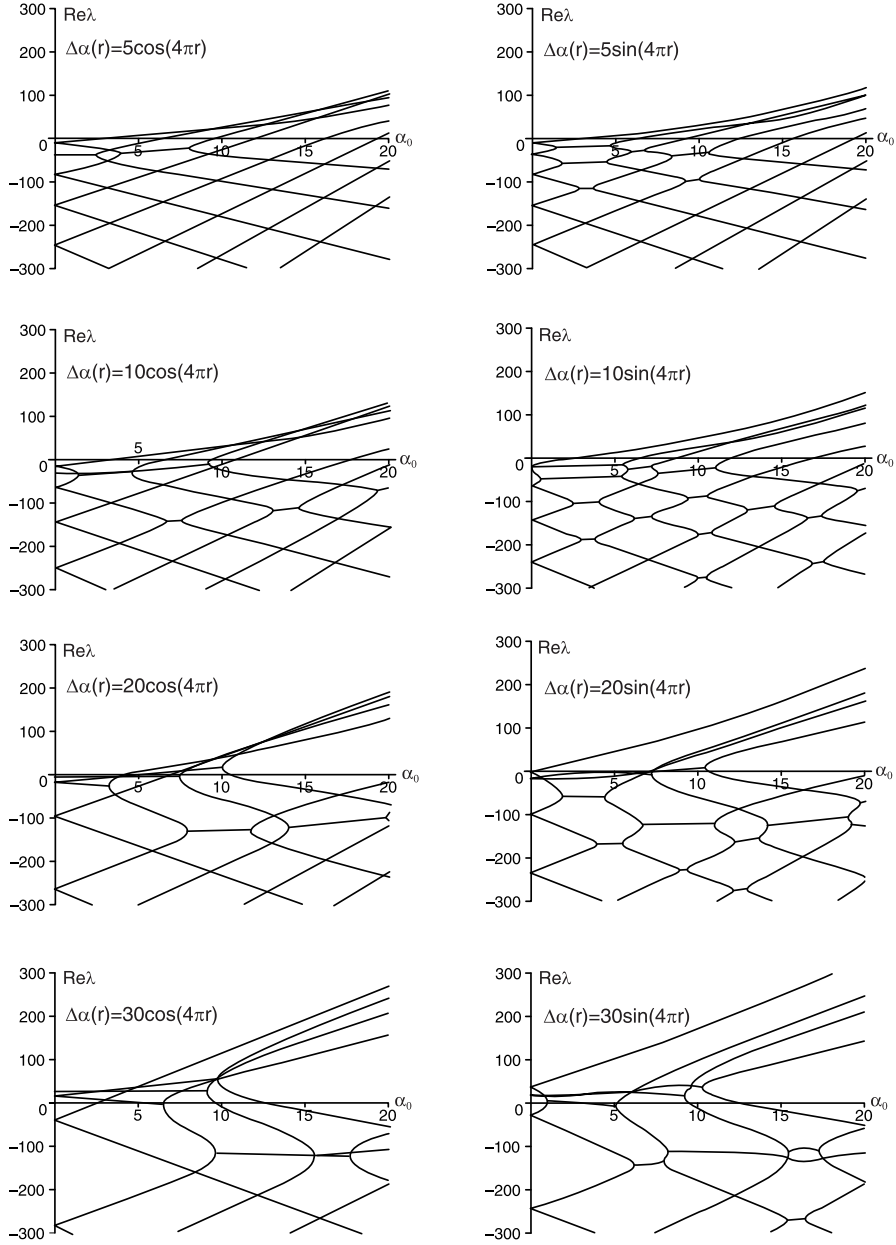}
    \end{center}
    \caption{Resonant deformation of the spectral mesh under strong
    and ultra-strong perturbations of pure harmonic type: $\Delta\alpha(r)=a_2\cos(4\pi r)$
      (left column) and $\Delta\alpha(r)=b_2\sin(4\pi r)$ (right
      column) with $a_2,b_2=5,10,20,30$. Strong perturbations lead not only
      to large spectral regions with complex conjugate eigenvalues,
      but also to a lifting of the corresponding complex branches
      into the upper $(\alpha_0,\Re \lambda)-$plane, i.e. they lead
      to overcritical dynamo regimes ($\Re \lambda>0$)
      of oscillatory type ($\Im \lambda\neq0$).}
    \label{fig7}
    \end{figure}

Up to now we used the Galerkin method for investigations of weak
perturbations over an $\alpha_0=$const background. In Fig.
\ref{fig7}  we demonstrate that the method works for strong
perturbations as well. We observe that, increasing the strength of
the pure harmonic perturbations from $a_2,b_2=5$ up to $a_2,b_2=30$,
the regions with complex conjugate spectral contributions grow and
finally intersect each other. Additionally, they shift into the
upper $(\alpha_0,\Re \lambda)-$plane leading to overcritical
oscillatory dynamo regimes $(\Re \lambda>0, \Im \lambda\neq 0)$. An
estimate of critical $\alpha-$profiles, for which such a transition
to the upper $(\alpha_0,\Re \lambda)-$half-plane starts to occur,
can be given within a first-order (linear) perturbative
approximation by assuming $\Re \lambda (\alpha_0)=0$ for the
exceptional point closest to the $(\Re \lambda=0)-$line. Relations
\rf{h16}, \rf{h15-1} and $\Re \lambda=\lambda_0^\nu+\Re
\lambda_1=\lambda_0^\nu+\pi M a_0/4$ yield this condition in terms
of the Fourier components $Q_j$ of such a critical $\alpha-$profile
as
\be{h15-3}
\Re \lambda=\frac{\pi^2}{4}\left(M^2-j^2\right)+\frac \pi 4 M
\left(1-\frac{M^2}{j^2}\right)^{1/2}|Q_j|=0.
\ee
This relation may be used for testing concrete $\alpha-$profiles on
their capability to produce complex eigenvalues in the right
spectral half-plane $(\Re \lambda>0, \Im \lambda \neq 0)$.

We restrict our present consideration of $(l=0)-$models to this
first numerical output and the analytical estimate, expecting
physically more relevant results from extending the present methods
to models with physically realistic boundary conditions. In the next
section we present some first few results on the $(l\ge 1)-$sectors
of the $\alpha^2-$dynamo model.

\section{Numerical techniques and examples for the $(l\ge 1)-$sectors\label{l>1}}

In this section, we reshape the general results of section
\ref{unfold} in a form suitable for numerical investigations and
demonstrate them on a first concrete model from the $(l=1)-$sector.

We start by representing the perturbation terms $[\fB
\fu_m^{\delta}, \fu_n^{\varepsilon}]$ from Eqs. \rf{10p1},
\rf{10p2} as
\ba{n10}
[\fB \fu_m^{\delta}, \fu_n^{\varepsilon}]
&=&\int_0^1\varphi\left[\left(\varepsilon\delta\sqrt{\rho_m\rho_n}+\frac{l(l+1)}{r^2}\right)u_m u_n+u_m'u_n'\right]dr\nn\\
&=&2(\rho_m\rho_n)^{1/4}\int_0^1\varphi(r) g_{mn}^l(r)dr, \quad m,n\in \ZZ^+
\label{10nO}
\ea
or equivalently
\ba{n10-1}\fl
[\fB \fv_m,\fv_n]&=&\int_0^1\varphi\left[\left(\varepsilon_m\varepsilon_n\sqrt{\rho_{|m|}\rho_{|n|}}
+\frac{l(l+1)}{r^2}\right)u_{|m|} u_{|n|}+u_{|m|}'u_{|n|}'\right]dr\nn\\
\fl &=&\int_0^1\varphi(r) g_{mn}^l(r)dr, \quad m,n\in \ZZ^*\,.
\ea
The functions $g_{mn}^l$ are symmetric $g_{mn}^l=g_{nm}^l$ and from
the Fr\'echet (functional) derivative\footnote{The Fr\'echet
derivative $\nabla f (x)$ of a function $f(x)$ over an open set
$X\ni x$ of a Banach space $F\supset X$ is defined as $
f(x+u)-f(x)=\nabla f(x)u +o(\|u\|) $
\cite{gradient1,gradient2,gradient3} what for the functional
$f(\varphi):=[\fB \fv_m, \fv_n]$ can be reshaped as $f(\varphi
+\chi)-f(\varphi)=(\nabla_\varphi f(\varphi),\chi)+o(\|\chi\|)$.}
\be{n10a} \frac{\delta [\fB \fv_m, \fv_n]}{\delta \varphi
(r)}\equiv \nabla_{\varphi}[\fB \fv_m, \fv_n]=g_{mn}^l(r)
\ee
we find that they can be naturally interpreted as components of the
perturbation gradient in the Krein space $\cK$. Their explicit
representation in terms of Bessel functions is given in
\ref{gradients}. Representation \rf{10nO} may prove especially
useful for the optimization of $\alpha$-profiles with regard to
given constraints or experimental requirements.

First, we note that in terms of the gradient functions $g_{mn}^l(r)$
the defining equation \rf{12p} for the first order spectral
perturbations $\lambda_1$ reduces to
\ba{n11}
&&\lambda_1^2- \lambda_1\int_0^1\varphi\left(\varepsilon g^l_{nn}+
\delta g^l_{mm}\right)dr+\nn\\
&&+\varepsilon\delta\left[\left(\int_0^1\varphi
g^l_{mm}dr\right)\left(\int_0^1\varphi
g^l_{nn}dr\right)-\left(\int_0^1\varphi g^l_{mn}dr
\right)^2\right]=0
\ea
with solutions
\ba{n12}
\lambda_{1,\pm}=\frac 1 2 \int_0^1\varphi \left(\varepsilon
g^l_{nn}+ \delta g^l_{mm}\right)dr \nn \\
\pm\frac 12 \sqrt{\left[\int_0^1\varphi \left( \varepsilon
g^l_{nn}- \delta g^l_{mm}\right)dr\right]^2+4\varepsilon\delta
\left[\int_0^1\varphi g^l_{mn}dr \right]^2}.
\ea

Comparison with the results for the hyper-idealized $s-$wave
sector (disk dynamo) shows that apart from the generic Krein space
related behavior (no complex eigenvalues for intersecting spectral
branches of the same type, $\delta=\varepsilon$, and possible
formation of complex eigenvalues for branches of different type,
$\varepsilon\neq \delta$) we find a generalization of the offset
and oscillation contributions for  $\varepsilon\neq \delta$ type
intersections. In rough analogy, the role of a transition
preventing offset is played by the 'diagonal' terms
\be{n13}
\int_0^1\varphi \left( g^l_{nn}+ g^l_{mm}\right)dr
\ee
whereas the 'off-diagonal' terms
\be{n14}
\left[\int_0^1\varphi g^l_{mn}dr \right]^2
\ee
enhance a possible transitions to complex eigenvalues, i.e. a
transition occurs for
\be{n15} \varepsilon\neq\delta\quad \cap\quad
\left(\frac{1}{2}\int_0^1\left(g^l_{nn}+g^l_{mm}\right)\varphi dr
\right)^2< \left(\int_0^1 g_{mn}^l\varphi dr\right)^2.
\ee
The condition \rf{n15} yields an explicit classification criterion
for $\alpha-$profile perturbations with regard to their capability
to induce complex eigenvalues. It may serve as an efficient search
tool for concrete $\alpha(r)-$profiles --- and transforms in this
way some general observations  of Ref. \cite{GS-jmp1} into a
technique of direct applicability.
\begin{figure}[htb]
\begin{center}
\includegraphics[angle=0, width=0.99\textwidth]{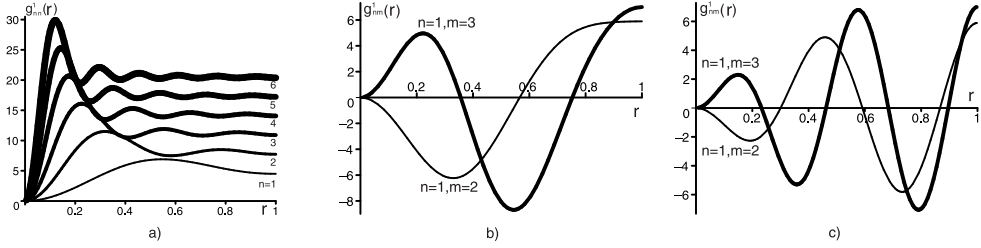}
\end{center}
\caption{Functions $g_{nm}^l(r)$ for $l=1$, $n=m$ (a) and $n=1$,
$m=2,3$ in the cases $\varepsilon\delta=1$ (b) and
$\varepsilon\delta=-1$ (c).} \label{fig4}
\end{figure}

\begin{figure}[htb]
\begin{center}
\includegraphics[angle=0, width=0.9\textwidth]{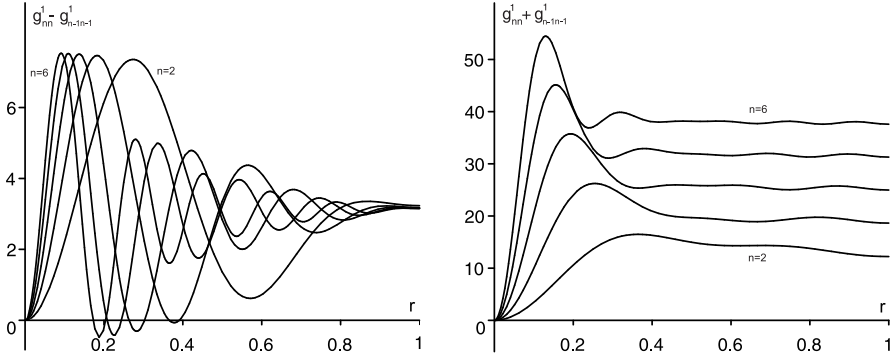}
\end{center}
\caption{Gradient function differences $g_{nn}^1(r)-g_{n-1\,
n-1}^1(r)$ as they enhance the real valued unfolding of diabolical
points in the upper $(\alpha_0,\Re \lambda)-$plane (left figure), as
well as offset gradient functions (integration kernels) of the type
$g_{nn}^1(r)+g_{n-1\, n-1}^1(r)$ for $n=2,\ldots, 6$ (right
figure).} \label{fig5}
\end{figure}

The functions $g_{nm}^l(r)$ for $l=1$ and different values of $n$,
$m$ and $\epsilon\delta$ are shown in Fig~\rf{fig2}. One clearly
sees that the 'diagonal' functions with $m=n$ are always
non-negative and the graphics of any two of them, $g_{nn}^1(r)$,
$g_{mm}^1(r)$ with $m\neq n$ intersect only marginally (see Fig.
\ref{fig5}). Their sums $g_{nn}^1(r)+g_{mm}^1(r)$ are strictly
sign-preserving functions so that they act as averaging integration
kernels. In contrast, the 'off-diagonal' functions $g_{nm}^1(r)$
with $m\ne n$ show strong sign changes and in this way they act as
filter kernels. Hence, the main qualitative roles of the 'diagonal'
and 'off-diagonal' functions as 'offset' and 'oscillation filter'
functions remain preserved also for the $(l=1)-$sector.
Their subtle interplay is crucial for the $\alpha$-profile to act as
generator of complex eigenvalues.
\begin{figure}[htb]
\begin{center}
\includegraphics[angle=0, width=0.5\textwidth]{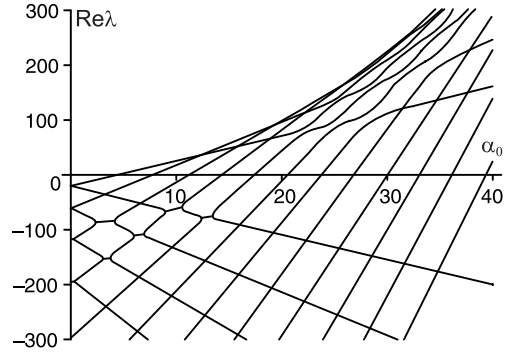}
\end{center}
\caption{Deformations of the spectral mesh with unfolding of
diabolical points for the $(l=1)-$sector of a model with
$\alpha(r)=\alpha_0+2.5 \cos(6\pi r)$.} \label{fig6}
\end{figure}

As in the $(l=0)-$sector,  spectral deformations over a certain
parameter space region can be studied numerically. For
$\alpha-$profiles $\alpha(r)=\alpha_0+\Delta\alpha(r)$ the
corresponding approximation matrix of the Galerkin method reads
simply
\be{n16}
A_{mn}[\alpha]=\lambda_m\delta_{mn}+\varepsilon_m
\int_0^1\Delta\alpha(r) g_{mn}^l(r)dr\,.
\ee
In Fig. \ref{fig6} we illustrate the method for an $\alpha-$profile
$\alpha(r)=\alpha_0+2.5 \cos(6\pi r)$ and a similar approximation
subspace $\cL=\vspan(\fv_{12},\ldots,\fv_{-12})\subset \cK$ as in
the previous section. For the $(l=1)-$mode Riccati-Bessel functions
a cosine perturbation is no longer an exact resonance mode and the
deformations of the mesh are spreading over a broader parabola-like
region.

Finally, we note that explicit analytical considerations of the
spectrum in the $(l>0)-$sector are obstructed by the lack of simple
transformation rules between Riccati-Bessel functions as well as of
simple expressions for integrals over triple products of spherical
Bessel functions in case of finite integration intervals.

\section{Conclusions and discussions}
In the present work, the spectral properties of spherically
symmetric MHD $\alpha^2-$dynamos with idealized boundary conditions
have been studied. Using the fundamental symmetry of the dynamo
operator matrix and the solution set of a model with constant
$\alpha-$profile, a Krein space related perturbation theory as well
as a Galerkin technique for numerical investigations have been
developed. As analytical result of the first-order perturbation
theory we found a strongly pronounced $\alpha-$resonance pattern in
the unfolding of diabolical points. The resonance behavior reflects
the correspondence between the characteristic length scale of
$\alpha-$perturbations and the decay rates of the coherently induced
field excitations. The observed correlations will strongly affect
the specifics of reversal processes of dynamo maintained magnetic
fields \cite{SG-prl,EPSL-1} and support corresponding numerical
simulations on more realistic dynamo setups \cite{giesecke-2}. For
the $(l=0)-$sector, a Fourier component based estimation technique
has been developed for obtaining the critical $\alpha-$profiles at
which the eigenvalues enter the right spectral half-plane with
non-vanishing imaginary components (at which overcritical
oscillatory dynamo regimes form). The analytical results on the
perturbative unfolding of diabolical points have been supplemented
by numerical studies of the deformations of the dynamo operator
spectrum. The capability of the used Galerkin approach has been
demonstrated in extending the strength of the $\alpha-$perturbations
from weakly perturbed regimes up to ultra-strong perturbations.
Extensions of the presented techniques to spherically symmetric
$\alpha^2-$dynamos with realistic boundary conditions as well as to
models of $\cP\cT-$symmetric quantum mechanics are straight forward.

\section*{Acknowledgements}
We thank  G. Gerbeth, H. Langer and C. Tretter for useful
discussions, and F. Stefani and M. Xu additionally for
cross-checking our Galerkin based results with other numerical
codes. The work has been supported by the German Research Foundation
DFG, grant GE 682/12-2, (U.G.) as well as by the CRDF-BRHE program
and the Alexander von Humboldt Foundation (O.N.K.).
\appendix

\section{Bessel function relations\label{bessel}}
The solutions $u_n(r)$ of the eigenvalue problem \rf{i9}
\be{a2}\fl
A_lu_n=\rho_n u_n,\qquad u_n(r\searrow 0)=u_n(r=1)=0,\qquad
A_l=-\partial_r^2+\frac{l(l+1)}{r^2}
\ee
are so called Riccati-Bessel functions \cite{abramowitz}
\be{a2a}
u_n(r)\sim \sqrt{\rho_n}r j_l(\sqrt{\rho_n}r)
\ee
where $j_l(\sqrt{\rho_n}r)$ denote spherical Bessel functions. They
can be expressed in terms of Bessel functions $J_{l+\frac
12}(\sqrt{\rho_n}r)$ and for definiteness we represent the solutions
$u_n$ as
\be{a1}
u_n(r)=N_n\sqrt r J_{l+\frac 12}(\sqrt{ \rho_n}r)
\ee
with $N_n$ a normalization coefficient.

The orthogonality of these solutions can be easily verified with the
help of well known Bessel function relations
\cite{abramowitz,Watson,Luke}. For $m\neq n$ it holds
\ba{a3}\fl
(u_m,u_n)=N_mN_n\int_0^1 rJ_{l+\frac 12}(\sqrt{
\rho_m}r)J_{l+\frac 12}(\sqrt{ \rho_n}r) dr\\ \fl
=\left.N_mN_n\frac{r}{\rho_m-\rho_n}\left[J_{l+\frac 12}(\sqrt{
\rho_m}r)\partial_r J_{l+\frac 12}(\sqrt{ \rho_n}r)-J_{l+\frac
12}(\sqrt{ \rho_n}r)\partial_r J_{l+\frac 12}(\sqrt{
\rho_m}r)\right]\right|^1_0\, .\nn
\ea
This expression vanishes because of $J_{l+\frac 12}(0)=J_{l+\frac
12}(\sqrt{ \rho_n})=0$ and $|\partial_r J_{l+\frac
12}(\sqrt{\rho_k}r)|_{r\searrow 0}<\infty $. In a similar way it
holds \cite{abramowitz,Watson,Luke}
\ba{a4}\fl
\|u_n\|^2=(u_n,u_n)=N_n^2\int_0^1 rJ_{l+\frac 12}^2(\sqrt{ \rho_n}r)
dr&=&\frac 12 N_n^2\left.\left[\partial_r J_{l+\frac
12}(\sqrt{ \rho_n}r)\right]^2\right|_{r=1}\nn\\
&=&\frac 12 N_n^2 J_{l+\frac 32}^2(\sqrt{ \rho_n})
\ea
so that $N_n=\sqrt 2/J_{l+\frac 32}(\sqrt{ \rho_n})$ gives
$(u_n,u_n)=1$.

\section{Explicit structure of the approximation matrix $\bA[\alpha]$\label{galerkin}}
In the $(l=0)-$sector, the relations \rf{h5}, \rf{h6}, \rf{h11a}
lead for the approximation matrix $\bA[\alpha]=\bfeta^{-1}(\tilde
A_{mn})_{m,n=-N}^N$, $\tilde A_{mn}=\left[\fA_\alpha
\fv_m,\fv_n\right]$ of \rf{d6} over an $\alpha-$profile
$\alpha(r)=\alpha_0+\Delta\alpha(r)$ to the following structure
\ba{g0}A_{mn}[\alpha]&=&\lambda_m\delta_{mn}+\varepsilon_m \pi
\sqrt{|mn|}\int_0^1\Delta\alpha(r)\cos[(m-n)\pi r] dr\nn\\
&=&\left[-(\pi m)^2+\alpha_0\pi m\right]\delta_{mn}+\varepsilon_m\frac \pi
2\sqrt{|mn|}Q_{m-n}\,.
\ea
In the case of  $\Delta\alpha(r)=a_k\cos (2\pi k r)$ this gives:
\be{g1}\fl
A_{mn}[\alpha_0,a_k]=\left[-(\pi m)^2+\alpha_0\pi m\right]\delta_{mn}+\varepsilon_m\frac \pi
2\sqrt{|mn|}a_k(\delta_{m,n+2k}+\delta_{m,n-2k})\, ,
\ee
i.e. $\bA[\alpha]$ has the eigenvalues $\lambda_m$ of the
unperturbed $\fA_{\alpha_0}$ on its diagonal and possesses two
subdiagonals with non-vanishing entries a distance $\pm2k$ aside the
main diagonal. For perturbations $\Delta\alpha(r)=b_k\sin (2\pi k
r)$ one finds
\ba{g2} A_{mn}[\alpha_0,b_k]&=&\left[-(\pi m)^2+\alpha_0\pi
m\right]\delta_{mn}+\nn\\
 &&+\varepsilon_m
\sqrt{|mn|}\, \left[1-(-1)^{m-n}\right]\frac{2kb_k}{4k^2-(m-n)^2}\,.
\ea
\section{Explicit expressions for the gradient functions $g_{mn}^l(r)$\label{gradients}}

Explicit expressions for the gradient components $g_{mn}^l(r)$ can
be derived from the representation \rf{i10} of the eigenfunctions
$u_n^{\varepsilon_n}$ and the defining relation \rf{n10}. As result
one obtains in $(n,\varepsilon_n)\in \ZZ^+\times \ZZ_2$ notations
\ba{b1} \fl
 && g_{mn}^l(r)=\frac{\left(\varepsilon_m\varepsilon_n\sqrt{\rho_m\rho_n} r^2+l^2+l\right)
  J_{l+1/2}\left(r\sqrt{\rho_m}\right)J_{l+1/2}\left( r\sqrt{\rho_n}\right)}
{r\left(\rho_m\rho_n\right)^{1/4}
J_{l-1/2}(\sqrt{\rho_m})J_{l-1/2}(\sqrt{\rho_n})}+ \nn \\
\fl
 &&+
  \frac{\left[r\sqrt{\rho_m}J_{l-1/2}\left(r\sqrt{\rho_m}\right)-l J_{l+1/2}\left(r\sqrt{\rho_m}\right)\right]
\left[ r\sqrt{\rho_n}J_{l-1/2}\left(r\sqrt{\rho_n}\right)-
  l J_{l+1/2}\left(r\sqrt{\rho_n}\right)\right]}
{r\left(\rho_m\rho_n\right)^{1/4}
J_{l-1/2}\left(\sqrt{\rho_m}\right)J_{l-1/2}\left(\sqrt{\rho_n}\right)}\nn\\
\fl \ea and $g_{mn}^l(r=1)=\left(\rho_m\rho_n\right)^{1/4}$.

\end{document}